\documentclass[pra,aps,twocolumn,nofootinbib,groupedaddress,10pt]{revtex4-2} %superscriptaddress

\usepackage{graphicx}
\usepackage[nointegrals]{wasysym} %
\usepackage[export]{adjustbox}
\usepackage{amsmath,amsfonts,amssymb,latexsym}
\usepackage{hhline}
\usepackage{bm}
\usepackage{verbatim}
\usepackage{enumitem}
\usepackage{mathrsfs}
\usepackage{slashed}
\usepackage{empheq}
\usepackage{physics}

\usepackage{graphicx}
\usepackage{dcolumn}
\usepackage{bm}
\usepackage{multirow}

\usepackage[normalem]{ulem}

\usepackage{amsmath}
\usepackage{amsthm}
\usepackage{amstext}
\usepackage{amssymb}
\usepackage{mathrsfs}
\usepackage{amsfonts}
\usepackage{amsbsy} 

\usepackage{braket}

\usepackage[all,matrix,cmtip]{xy}
\usepackage{mathtools}

\usepackage{xcolor}
\definecolor{red}{rgb}{1,0,0}
\definecolor{blue}{rgb}{0,0,1}
\definecolor{dblue}{rgb}{0,0,0.4}
\definecolor{green}{rgb}{0,1,0}
\definecolor{black}{rgb}{0,0,0}
\definecolor{white}{rgb}{1,1,1}
\definecolor{pastelblue}{RGB}{20,93,160}

\definecolor{brn}{rgb}{.8,.4,.0}
\definecolor{redo}{rgb}{1,.5,.0}
\definecolor{ddgrn}{rgb}{0,0.4,0}
\definecolor{dgrn}{rgb}{0,0.55,0}
\definecolor{dbl}{rgb}{0,0,0.5}

\usepackage[colorlinks,citecolor=pastelblue,linkcolor=pastelblue,urlcolor=pastelblue]{hyperref}

\renewcommand{\Im}{{\rm Im}}

\newcommand{\bpm}{\begin{pmatrix}}
	\newcommand{\epm}{\end{pmatrix}}
\newcommand{\bmm}{\begin{matrix}}
	\newcommand{\emm}{\end{matrix}}
\newcommand{\bvm}{\begin{vmatrix}}
	\newcommand{\evm}{\end{vmatrix}}

\usepackage{euscript}

\makeatletter
\newsavebox{\@brx}
\newcommand{\llangle}[1][]{\savebox{\@brx}{\(\m@th{#1\langle}\)}%
	\mathopen{\copy\@brx\kern-0.5\wd\@brx\usebox{\@brx}}}
\newcommand{\rrangle}[1][]{\savebox{\@brx}{\(\m@th{#1\rangle}\)}%
	\mathclose{\copy\@brx\kern-0.5\wd\@brx\usebox{\@brx}}}
\makeatother

\newcommand{\bs}{\boldsymbol}

\allowdisplaybreaks

\usepackage{tikz}
\definecolor{myRed}{RGB}{188,0,4} % #BC0004
\definecolor{myGray}{RGB}{146,146,146} % #929292
\definecolor{myBlue}{RGB}{0,0,133} % #000085

\begin{document}

%\preprint{APS/123-QED}

\title{Anyon delocalization transitions out of a disordered FQAH insulator}% Force line breaks with \\

\author{Zhengyan Darius Shi}%
\email{zdshi@mit.edu}
\affiliation{
Department of Physics, Massachusetts Institute of Technology,
Cambridge, Massachusetts 02139, USA
}

\author{T. Senthil}
\email{senthil@mit.edu}
 %\homepage{}
\affiliation{
Department of Physics, Massachusetts Institute of Technology,
Cambridge, Massachusetts 02139, USA
}%

\date{\today}

\begin{abstract}
Motivated by the experimental discovery of the fractional quantum anomalous Hall (FQAH) effect, we develop a theory of doping-induced transitions out of the $\nu = 2/3$ lattice Jain state in the presence of quenched disorder. We show that disorder strongly affects the evolution into the conducting phases described in our previous work. The delocalization of charge $2/3$ anyons leads to a chiral topological superconductor through a direct second order transition for a smooth random potential with long-wavelength modulations. The longitudinal resistance has a universal peak at the associated quantum critical point. Close to the transition, we show that the superconducting ground state is an ``Anomalous Vortex Glass (AVG)'' stabilized in the absence of an external magnetic field. For short-wavelength disorder, this transition generically splits into three distinct ones with intermediate insulating topological phases.  If instead, the charge $1/3$ anyon delocalizes, then at low doping the result is a Reentrant Integer Quantum Hall state with $\rho_{xy} = h/e^2$. At higher doping this undergoes  a second transition to a Fermi liquid metal. We show that this framework provides a plausible explanation for the complex phase diagram recently observed in twisted MoTe$_2$ near $\nu = 2/3$ and discuss future experiments that can test our theory in more detail.

\end{abstract}

\maketitle

The discovery of zero-field fractional quantum anomalous Hall (FQAH) phases in twisted MoTe$_2$ (\textit{t}MoTe$_2$) and rhombohedral graphene~\cite{Park2023_FQAHTMD,Zeng2023_FQAHTMD,Cai2023_FQAHTMD,Xu2023_FQAHTMD,Lu2023_FQAHPenta,Lu2025_EQAH} opens up a new frontier in topological materials research. In common with conventional fractional quantum Hall (FQH) phases stabilized in a large external magnetic field~\cite{Stormer1999_FQHreview}, FQAH phases host anyon excitations with fractional charge and fractional statistics~\cite{Leinaas1977_anyon,Wilczek1982_anyon1,Wilczek1982_anyon2,Laughlin1983_FQHtheory,Halperin1984_anyonFQH}. However, unlike in the large-field setting, anyons in FQAH phases are not confined to cyclotron orbits and carry a non-trivial band dispersion.  Moreover, the two-dimensional Van der Waals materials that realize FQAH phases come with additional tuning knobs such as the displacement field, which allow for a continuous modulation of the anyon dispersion. This tunability enables us to explore the interplay between topology, electron kinetic energy and interactions. 

The existence of non-trivial anyon dispersion has a profound impact on the fate of FQAH phases upon doping. In the large-field setting, anyons doped into the system carry no kinetic energy and become localized in the presence of disorder (see ~\cite{Huckestein1995_disorder_QHreview} and references therein). This mechanism explains the formation of quantum Hall plateaus which persist until the basic physics responsible for the emergence of anyons is destroyed. In contrast, the kinetic energy of anyons doped into an FQAH insulator  
competes with the disorder-induced tendency towards localization. This competition decreases the width of plateau regions and favors the formation of itinerant anyonic quantum phases upon exiting the plateau.

Recent theoretical works have explored the doped FQAH regime in an ideal limit with no disorder~\cite{Shi2024_doping,Kim2024_chiral_anyonSC,Shi2025_doping_nonabelian}, uncovering a variety of interesting phases including topological superconductors\footnote{The notion of superconductivity developing in a mobile gas of anyons - dubbed ``anyon superconductivity" - is an old idea~\cite{Laughlin1988_anyonSC,Fetter1989_anyonSC_RPA,Chen1989_anyonSC,Lee1989_anyonSC} that has been refined and extended recently~\cite{Shi2024_doping,Divic2024_anyonSC}. We caution that the superconducting state itself has universal properties that are the same as an electronic superconductor, and need not have any anyon excitations. Thus a better terminology is to refer to the superconductor stabilized in a mobile anyon fluid as ``anyon-induced superconductivity".} and charge-ordered Fermi (and non-Fermi) liquids. Subsequently, an experiment in \textit{t}MoTe$_2$ indeed finds a narrow plateau around the Jain state at lattice filling $\nu = 2/3$ and a superconductor that emerges at $\nu > 2/3$ outside the plateau region~\cite{Xu2025_SCdopeTMD}. On the opposite side $\nu < 2/3$, a reentrant integer quantum anomalous Hall (RIQAH) state apparently arises in between the plateaus surrounding $\nu = 2/3$ and $\nu = 3/5$. This rich phenomenology contrasts with the sequence of FQH plateau transitions that occur in the conventional quantum Hall setting within the same range of Landau level filling~\cite{willett1987observation,Stormer1999_FQHreview}.

Motivated by these experimental discoveries, we specialize to the lattice Jain state at filling $\nu = 2/3$ and study its evolution into the conducting states described in our previous work~\cite{Shi2024_doping}. First, we study the chemical potential tuned FQAH-SC phase transition in both the clean and dirty limits. We also discuss a bandwidth tuned FQAH-SC transition at fixed lattice filling $\nu = 2/3$. Our analysis of the dirty limit shows that quenched disorder modifies the FQAH-SC evolution in striking ways that we elaborate on below. Second, we also analyse the chemical potential tuned evolution from the FQAH state to a charge-ordered Fermi liquid metal (also obtained in our previous work). In the presence of disorder, while the details of the evolution are complicated, the charge-ordered metal is replaced by an integer quantum Hall state in a portion of the phase diagram. 

\begin{figure}[!ht]
    \centering
    \includegraphics[width=0.8\linewidth]{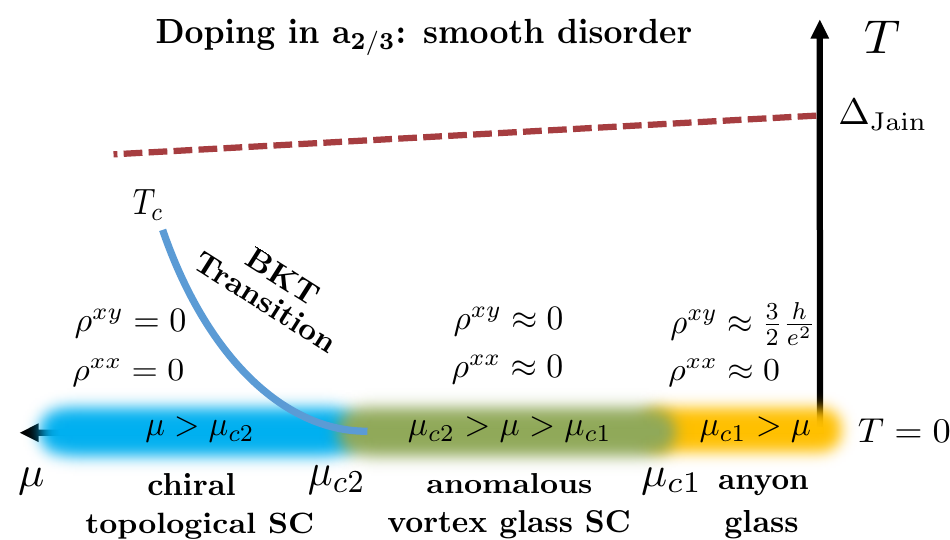}
    \caption{Onset of $T_c$ and $\rho_s$ as a function of $\mu$ across the FQAH-SC evolution with smooth disorder. In this phase diagram, we assume that $\mu$ is close enough to the chemical potential of the parent FQAH insulator so that the Jain gap $\Delta_{\rm Jain}$ remains much higher than $T_c$. The anomalous vortex glass (AVG) SC and the chiral topological SC both have zero resistance at $T = 0$. At $T > 0$, the AVG has a small nonzero resistance due to the creep motion of vortices. As a result, there is no sharp finite-temperature phase transition into the AVG and $T_c = 0$ for $\mu_{c2} > \mu > \mu_{c1}$.}
    \label{fig:FQAH_SC_smooth_rhos}
\end{figure}

\begin{figure}[!ht]
    \centering
    \includegraphics[width=\linewidth]{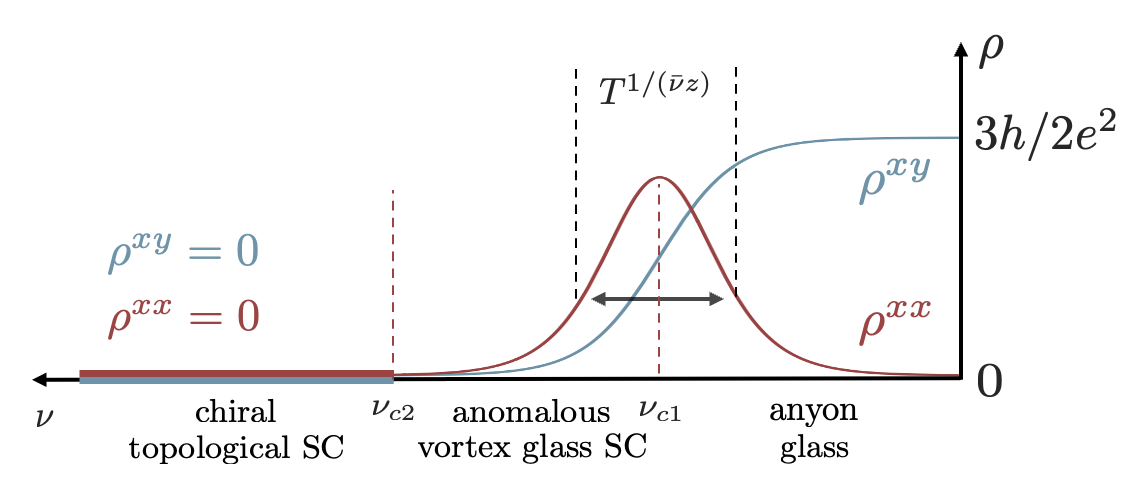}
    \caption{Evolution of $\rho$ upon doping $a_{2/3}$ at fixed $T > 0$ with smooth disorder. To match the experimental convention, our plot shows $\nu$ increasing towards the left. $\nu_{c1}$ marks the $T=0$ critical point with universal transport and $\nu_{c2}$ defines the transition from the AVG to the chiral topological SC. Whether the chiral topological SC can be observed in experiment depends on the width of the AVG region, which requires a more detailed microscopic calculation.} 
    \label{fig:FQAH_SC_smooth_rho}
\end{figure}

\begin{figure}
    \centering
    \includegraphics[width=\linewidth]{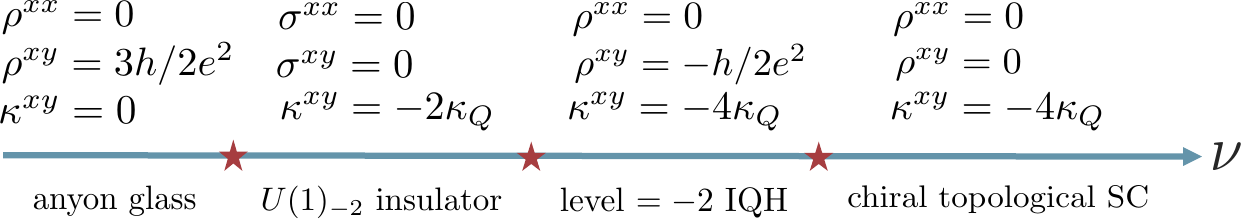}
    \caption{With short-wavelength disorder, the FQAH-SC evolution passes through three intermediate phases with quantized $\rho$ and $\kappa^{xy}$. In the phase diagram above, $U(1)_{-2}$ is a neutral topological order and $\kappa_Q$ denotes the thermal Hall quantum corresponding to a single Majorana edge mode. }%\ts{Change $\rho^{xy}$ in the IQH state}
    \label{fig:FQAH_SC_rough}
\end{figure}

The main characters in our analysis are a pair of basic anyonic excitations of the Jain state, $a_{1/3}$ and $a_{2/3}$, which carry fractional charges $e/3$ and $2e/3$ respectively. The possible doping-induced phases as well as the universality class of phase transitions between them depend sensitively on the energetics of $a_{1/3}$ and $a_{2/3}$ in the undoped Jain state. When the energy gap of $a_{2/3}$ is smaller than twice the energy gap of $a_{1/3}$, the doped charges enter as a fluid of $a_{2/3}$ anyons.\footnote{See Refs.~\cite{Xu2025_anyon_bunching,gattu2025molecularanyonsfractionalquantum} for very recent calculations demonstrating bunching of anyons in a Landau level.} At low dopant density, disorder turns this fluid into a localized anyon glass. For a smooth disorder potential with only long-wavelength modulations, increasing doping induces a direct continuous transition into a chiral topological superconductor with $c_- = -2$ (corresponding to four chiral Majorana edge modes). The longitudinal resistance has a universal peak of $\mathcal{O}(h/e^2)$ at the $T = 0$ quantum critical point, which is broadened at non-zero temperatures (see Fig.~\ref{fig:FQAH_SC_smooth_rhos} and Fig.~\ref{fig:FQAH_SC_smooth_rho}). We argue that localized anyons in the anyon glass phase evolve into an ``Anomalous Vortex Glass" (AVG) in the superconductor sufficiently close to the critical point. The finite-temperature dynamics of this phase share many features with the field-induced vortex glass~\cite{fisher1989vortex,fisher1991thermal} in conventional type-II superconductors. In particular, vortex creep destabilizes the AVG SC at $T > 0$, resulting in a finite resistivity and non-linear current-voltage characteristics at low current densities. With short-wavelength disorder (e.g. from point defects), a direct FQAH-SC transition is forbidden and the system instead passes through two intermediate phases with quantized electrical and thermal transport (see Fig.~\ref{fig:FQAH_SC_rough}). The critical points between these phases have similarities with plateau transitions between FQH phases that we will comment on below. 

\begin{figure}
    \centering
    \includegraphics[width=\linewidth]{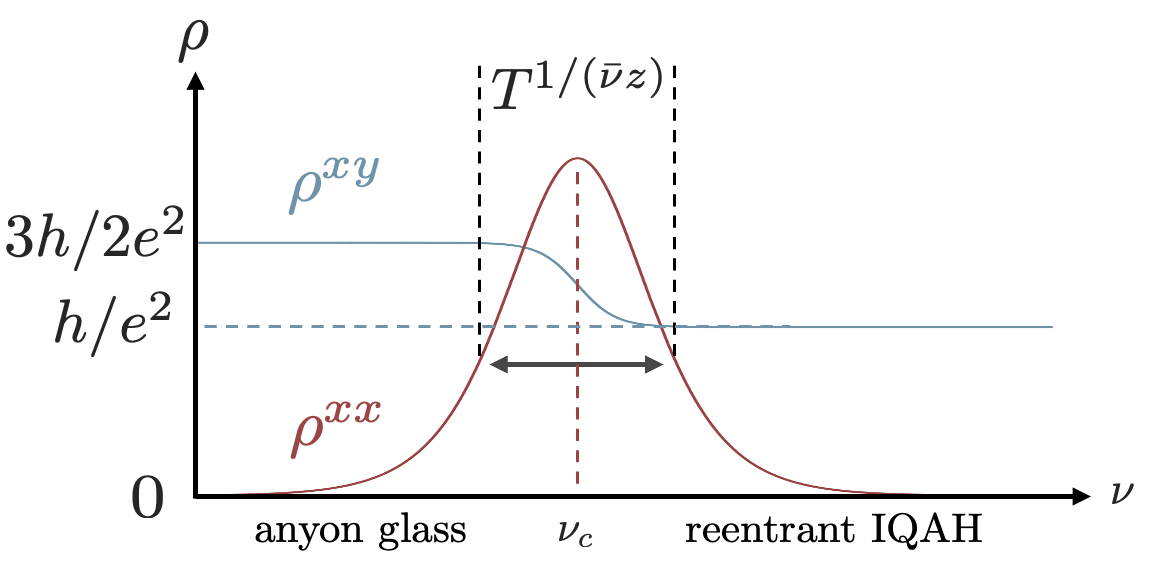}
    \caption{Evolution of $\rho^{xx}$ and $\rho^{xy}$ upon doping the $a_{1/3}$ anyon. At $\nu > \nu_c$, the inclusion of disorder turns the charge-ordered Fermi liquid in the clean model into a reentrant IQAH state.}
    \label{fig:FQAH_RIQAH_rho}
\end{figure}

When the energy gap of a localized $a_{2/3}$ anyon is larger than twice the energy gap of $a_{1/3}$, the dopant charges form a fluid of $a_{1/3}$ anyons instead. In this case, with both short and long-wavelength disorder, we argue that there is a direct transition from the anyon glass at low doping to an integer quantum Hall (IQH) state with $\sigma^{xy} = e^2/h$ at higher doping (see Fig.~\ref{fig:FQAH_RIQAH_rho}). 

The theory developed above applies to any doped lattice $\nu = 2/3$ Jain state. The main input from microscopics is the relative gap sizes of distinct anyons which determine whether $a_{1/3}$ or $a_{2/3}$ is doped into the system. Armed with our theoretical analysis, we discuss the observation of SC and RIQAH states proximate to the $\nu = 2/3$ FQAH state in \textit{t}MoTe$_2$.
 We propose an interpretation of the \textit{t}MoTe$_2$ phase diagram from Ref.~\cite{Xu2025_SCdopeTMD}, in which doping towards $|\nu| > 2/3$ induces a fluid of $a_{2/3}$ anyons, while doping towards $|\nu| < 2/3$ induces a fluid of $\bar a_{1/3}$ anyonic holes. This asymmetry in the particle and hole excitation spectrum leads to an asymmetric phase diagram in which $\nu > 2/3$ realizes a superconductor while $\nu < 2/3$ realizes a localized IQH state.

\paragraph*{Low energy theory near $\nu = 2/3$:} 

Following Ref.~\cite{Shi2024_doping}, we describe the Jain state at lattice filling $\nu = 2/3$ through a parton construction in which the microscopic electron $c$ is fractionalized as $c = f_1 f_2 f_3$. The low energy Lagrangian takes the form
\begin{equation}
    L = L[f_1, a] + L[f_2, b-a] + L[f_3, A-b] \,,
\end{equation}
where $a,b$ are emergent $U(1)$ gauge fields introduced to remove the redundancy in the parton decomposition. The equations of motion for $a_t$ and $b_t$ enforce the constraint $\rho_1 = \rho_2 = \rho_3 = \rho_c$, where $\rho_i/\rho_c$ is the charge density of $f_i$/$c$ respectively. In units where the lattice constant is set to 1, we can consider a uniform mean-field ansatz in which 
\begin{equation}
    \frac{\ev{\nabla \times \bs{a}}}{2\pi} = - \frac{1}{3} \,, \quad \frac{\ev{\nabla \times \bs{b}}}{2\pi} = \frac{1}{3} \,. 
\end{equation}
This choice ensures that the flux seen by $f_i$ is commensurate with its lattice filling $\nu_i = 2/3$. Therefore, each $f_i$ can form a Chern insulator with total Chern number $C_i$ determined by the detailed band structure. To construct the Jain state at $\nu = 2/3$, the unique Chern number assignment (up to permutations) is $C_1 = -2, C_2 = C_3 = 1$~\cite{Shi2024_doping}. With this choice, we can integrate out the gapped partons $f_i$ as well as the gauge field $a$ to obtain\footnote{We include in  the continuum effective theory a coupling to a background space-time metric $g$.  The gravitational Chern-Simons term $2\mathrm{CS}_g$ enables us to keep track of the thermal Hall conductance of various phases.}
\begin{equation}\label{eq:Jain_TQFT}
    L = \frac{3}{4\pi} bdb - \frac{1}{2\pi} b d A + \frac{1}{4\pi} A d A + 2 \mathrm{CS}_g \,. 
\end{equation}

Upon doping away from $\nu = 2/3$, the excess charges form a dilute itinerant anyon fluid. The key observation in Ref.~\cite{Shi2024_doping} is that the fate of this anyon fluid at zero temperature depends on the energetics of different anyons in the parent FQAH insulator. In the parton description, the $a_{1/3}$ anyon is sourced by $f_1$ (the zero-field analogue of the composite fermion) and the $a_{2/3}$ anyon is sourced by $f_2$ or $f_3$. Therefore, the question of which anyon gets doped into the system translates to the question of which parton band is closest to the chemical potential. We now consider both possibilities (doping in $a_{2/3}$ or $a_{1/3}$) and analyze the evolution out of the FQAH phase with and without quenched disorder.

\paragraph*{Doping the $a_{2/3}$ anyon: clean limit}

\begin{figure}
    \centering
    \includegraphics[width=0.8\linewidth]{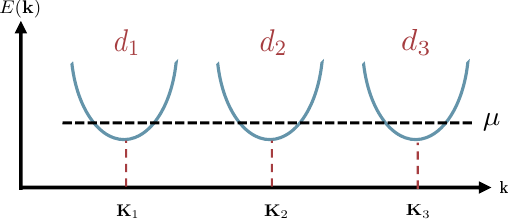}
    \caption{A schematic depiction of the continuum fields $d_I$ that create excitations of the $f_3$ band near its minima $\bs{K}_1, \bs{K}_2, \bs{K}_3$. When $\mu$ is above the band bottom, the $d_I$ particles see an emergent magnetic field proportional to their density, which is not shown in the figure.}
    \label{fig:d_bands}
\end{figure}
Let us first study the situation in which the $a_{2/3}$ anyon is doped into the system.  
In the absence of disorder, the phase transition out of the FQAH state occurs (within the parton description) when the chemical potential $\mu$ touches the minima of the unfilled $f_3$ band. Lattice translations $T_x, T_y$ act projectively on $f_3$ as it sees $2\pi/3$ flux per plaquette.  
Thus $f_3$ carries well-defined crystal momenta in a three-fold reduced Brillouin zone, and  
its band structure has three degenerate minima at $\bs{K}_1, \bs{K}_2, \bs{K}_3$ such that $|\bs{K}_I - \bs{K}_J|$ is a multiple of $2\pi/3$. Near the critical point, it is convenient to introduce three species of continuum fermionic fields $d_I$ that create $f_3$ excitations near $\bs{K}_I$ and transform projectively under lattice translation (see Fig.~\ref{fig:d_bands} for an illustration) 
\begin{equation}\label{eq:d_Iproj}
    T_x: d_I \rightarrow d_{I+1} \,, \quad T_y: d_I \rightarrow e^{2\pi i I/3} d_I \,. 
\end{equation}
In terms of these fields, the low energy effective Lagrangian takes the form
\begin{equation}\label{eq:L_clean_a2/3}
    \begin{aligned}
    &L_{\rm clean, a_{2/3}} \\
    &= \sum_{I=1}^3 d_I^{\dagger} \left[i \partial_t + \mu + A_t - b_t - \frac{(i\nabla + \bs{A} - \bs{b})^2}{2m}\right] d_I \\
    &\hspace{0.2cm} + L_{\rm int}[d_I] + \frac{3}{4\pi} b db - \frac{1}{2\pi} b d A + \frac{1}{4\pi} A d A + 2 \mathrm{CS}_g \,,
    \end{aligned}
\end{equation}
where $m$ is the effective mass of $f_3$ near the band minima and $L_{\rm int}$ includes repulsive interactions between the $d_I$ fermions. When $\mu < 0$, the $d_I$ fermions are gapped and we recover the Jain state described by \eqref{eq:Jain_TQFT}. When $\mu > 0$, the equations of motion for $b_t$ enforce the constraint $\sum_{I=1}^3 \rho_I = \frac{3}{2\pi} \ev{\nabla \times \bs{b}}$. Since $d_I$ feels an effective magnetic field $-\nabla \times \bs{b}$, this constraint implies that each $d_I$ is at Landau level filling $-1$ and forms an IQH state. Integrating out the filled Landau levels of $d_I$ gives the chiral topological charge-2 superconductor
\begin{equation}\label{eq:L_SCclean}
    L_{\rm SC} = \frac{2}{2\pi} b d A - \frac{2}{4\pi} A d A - 4 \mathrm{CS}_g \,. 
\end{equation}

It is natural to wonder how the expected $h/2e$ flux quantization works out in this superconductor given its origins from a  fluid of charge-$2/3$ anyons. We show in Appendix~\ref{app:AVG} that if we start with an annulus in the FQAH phase with an $a_{2/3}$ anyon trapped in the inner hole, then upon transitioning to the SC, a vorticity of $-2\pi \mod 6\pi$ in the phase of the Cooper pair order parameter is induced (corresponding to quantized flux $-h/2e \mod 3h/2e$). If instead an $a_{1/3}$ anyon is initially trapped in the inner hole, then a Cooper pair vorticity of $+2\pi \mod 6\pi$ is induced in the SC. The general phenomenon of anyons transmuting into vortices is well-known~\cite{senthil2000z,senthil2001fractionalization} in SC arising in, e.g., doped $Z_2$ spin liquid Mott insulators though the details, such as the $6\pi$ periodicity and the asymmetry between $\pm 2\pi$ vortices, are different in the present context.

At the critical point $\mu = 0$, the mean-field propagator for $d_I$ takes the simple form 
\begin{equation}
    G^{(d)}_{IJ}(\bs{k}, i\omega) = \frac{\delta_{IJ}}{i\omega - k^2/(2m)} \,.  
\end{equation}
As a result, the dynamical critical exponent is $z = 2$ and the correlation length exponent is $\bar{\nu} = 1/2$.  
A short-range interaction of the form $V_{IJ} \rho_I \rho_J$ is marginal at tree level. Following the same calculation as in the Bose-Hubbard model~\cite{fisher1989boson}, we can show that $V_{IJ}$ is marginally irrelevant and can be neglected at the IR fixed point. 

The interaction between $d_I$ and the gauge field $b$ is also marginal by power counting. Nevertheless, we can argue that the gauge and matter sectors decouple at zero temperature. The physical reason is that the dispersion of $d_I$ contains a ``particle" branch above $\mu$ but no hole branch below $\mu$. This means that the ground state contains no occupied fermionic modes and particle-hole excitations do not exist at zero temperature. Since the gauge field $b$ couples to particle-hole excitations, this observation immediately implies the decoupling between matter and gauge sectors at $T = 0$. On a more technical level, this is equivalent to the statement that poles of $G^{(d)}_{IJ}(\bs{k}, i\omega)$ in the complex $\omega$-plane always satisfy $\Im[\omega] < 0$. This pole structure guarantees that all loop corrections to the gauge field self-energy vanish identically. 

\paragraph*{Doping the $a_{2/3}$ anyon: dirty limit}  
For real experimental systems, quenched disorder is ubiquitous and modifies the above picture dramatically. The most relevant form of disorder near the phase transition is a generalized random potential that breaks the lattice translation symmetry completely
\begin{equation}
    L_{\rm dis} = W_{IJ}(\bs{r}) d^{\dagger}_I(\bs{r},t) d_J(\bs{r},t) + W_b(\bs{r}) \frac{\nabla \times \bs{b}(\bs{r}, t)}{2\pi} \,. 
\end{equation}
When $\mu$ is positive but small, each of the $d_I$ fermions wants to form an IQH state with a small energy gap $\omega_c(\delta) \sim \delta/m$, where $\delta \ll 1$ is the density of $d_I$. Let $\overline{W}_{IJ}, \overline{W}_b$ characterize the strength of the disorder potentials $W_{IJ}(\bs{r}), W_b(\bs{r})$. If $\omega_c(\delta) \ll \overline{W}_{IJ}, \overline{W}_b$, the mean-field quantum Hall state for $d_I$ is replaced by a localized state with vanishing $\sigma^{xx}$ and $\sigma^{xy}$. Gauge fluctuations turn this mean-field state into a localized ``anyon glass", leading to a plateau surrounding the $\nu = 2/3$ state with quantized $\sigma^{xy} = 2e^2/3h$. In the opposite limit where $\omega_c(\delta) \gg \overline{W}_{IJ}, \overline{W}_b$, the disorder potential is a weak perturbation to the mean-field IQH state of $d_I$ and we recover a disordered superconductor. 

The evolution from the anyon glass to the superconductor depends on the spatial profile of the disorder potentials. When $W_{IJ}(\bs{r})/W_b(\bs{r})$ is a smooth potential with only long-wavelength modulations, we can truncate the Fourier modes $W_{IJ}(\bs{q})/W_b(\bs{q})$ at some small momentum cutoff $q_{\rm max} \ll 1$ (in units where the lattice spacing is 1). In the low energy effective theory, this means that disorder potentials that transform under the action of lattice translations should be excluded. From \eqref{eq:d_Iproj}, it is clear that  
$d^\dagger_I T_{IJ} d_J$, with $T_{IJ}$ any $SU(3)$ matrix, transforms non-trivially under $T_x$ and/or $T_y$. Therefore, the only terms compatible with smooth disorder take the form
\begin{equation}\label{eq:Ldis_smooth}
    \begin{aligned}
    L_{\rm dis, smooth} &= W_d(\bs{r}) \sum_{I=1}^3 d^{\dagger}_I d_I + W_b(\bs{r}) \, \frac{\nabla \times \bs{b}(\bs{r})}{2\pi} \,.
    \end{aligned}
\end{equation}
\begin{comment}
Using the gauge constraint $\sum_I d^{\dagger}_I d_I = \frac{3}{2\pi} \nabla \times \bs{b}$, we can further simplify this term to
\begin{equation}\label{eq:Ldis_smooth}
    L_{\rm dis, smooth} = V(\bs{r}) \sum_{I=1}^3 d^{\dagger}_I d_I \,, 
\end{equation}
where $V(\bs{r}) = W_d(\bs{r}) + W_b(\bs{r})/3$. \end{comment} 
When $L_{\rm dis, smooth}$ is added to the critical theory, the combined Lagrangian enjoys an emergent $SU(3)$ symmetry\footnote{More precisely, the symmetry is $PSU(3) = SU(3)/\mathbb{Z}_3$.} that rotates the fermionic fields $d_I$. As we increase the dopant density $\delta$ (by tuning the chemical potential $\mu$), there is a critical $\delta_{c1}$ ($\mu_{c1}$) at which $d_I$ goes through an $SU(3)$-symmetric plateau transition with a jump in the Hall conductance $\Delta \sigma^{xy}_d = -3 e^2/h$. With gauge fluctuations, this turns into a direct transition from the anyon glass (with quantized $\sigma^{xy} = 2e^2/3h$) to the topological superconductor described by \eqref{eq:L_SCclean}. 

Let us discuss the anyon glass to SC transition in more detail, going beyond the lens of the specific parton description discussed above. On general grounds, this transition is expected to be continuous and the superconducting order is weak near the critical point. As we argue in Appendix~\ref{app:AVG}, sufficiently close to the transition, the localized anyons in the anyon glass phase evolve into a non-zero density of randomly pinned vortices of strength either $-h/2e$ or $h/e$ in the superconductor, with vanishing \textit{total} vorticity. Since these vortices emerge at zero external magnetic field and exhibit glassy dynamics~\cite{fisher1989vortex,fisher1991thermal}, we refer to this many-body state as an ``Anomalous Vortex Glass" (AVG) superconductor. This phase is characterized by a non-vanishing Edwards-Anderson superconducting order parameter. Like the field-induced vortex glass, the AVG is not stable to thermal fluctuations at $T > 0$ due to the creep motion of vortices. As a result, the linear resistance is nonzero at any $T > 0$. Furthermore the current-voltage characteristics exhibit non-linearities at a low current density scale $J_T \sim T^{1 + \frac{1}{|\theta|)}}$ with $\theta$ a universal exponent estimated~\cite{fisher1991vortex} to be around $0.5$. There is thus no sharp finite-temperature phase transition at small $\mu - \mu_{c1}$. Upon increasing $\mu$, the density of vortices gradually decreases and vanishes at a second critical point $\mu_{c2} > \mu_{c1}$. Dissipationless transport is restored at $T > 0$ for $\mu > \mu_{c2}$ and the finite-temperature transition out of this SC state belongs to the standard Berezinsky-Kosterlitz-Thouless (BKT) universality class.

While the precise critical exponents of the transitions at $\mu_{c1}$ and $\mu_{c2}$ are difficult to access analytically, there are a number of qualitative features that can be immediately deduced. Within the parton description,  the $d_I$ fermions are gapless but not gauge-invariant. The gauge-invariant electron operator $c$ can be constructed by combining $d_I$ with the gapped partons $f_1$ and $f_2$. Therefore, the microscopic electron remains gapped across the FQAH-SC evolution and single particle tunneling sees no spectral weight at frequencies below the $f_1/f_2$ gap. This may also be understood more generally as, by assumption, the $a_{1/3}$ gap stays non-zero across the evolution into the SC, and it is not possible to build the electron by fusing powers of $a_{2/3}$ alone.  

In terms of transport, general scaling arguments imply that the critical theory at $\mu_{c1}$ has a universal non-zero resistivity tensor $\rho_c$ with both longitudinal and Hall components of $\mathcal{O}(h/e^2)$~\cite{fisher1989boson,Cha1991_crit_transport_SIT,Damle1997_crit_transport,Sachdev1998_crit_transport}. This may be approximately related to the conductivity tensor $\sigma_d$ of $d_I$ through the Ioffe-Larkin rule  
\begin{equation}
    \rho_c = \begin{pmatrix}
        0 & -1/2 \\ 1/2 & 0 
    \end{pmatrix} + \left[\begin{pmatrix}
        0 & 1 \\ -1 & 0
    \end{pmatrix} + \sigma_d\right]^{-1} \,.
\end{equation}
Computing the inverse explicitly gives 
\begin{equation}
    \begin{aligned}
    \rho_c^{xx} &= \frac{\sigma_d^{xx}}{(\sigma_d^{xx})^2 + (1 + \sigma_d^{xy})^2} \,, \\
    \rho_c^{yx} &= \frac{1}{2} + \frac{1 + \sigma_d^{xy}}{(\sigma_d^{xx})^2 + (1 + \sigma_d^{xy})^2} \,. 
    \end{aligned}
\end{equation}

Thus the critical point is signaled by a universal peak in $\rho^{xx}$. At $T > 0$, the peak is broadened by thermal fluctuations, with a width that scales as $T^{1/(\bar \nu z)}$. The critical exponents $\bar \nu, z$ are constrained by standard arguments~\cite{fisher1989boson,fisher1990presence} for disordered superfluid-insulator transitions. With short-range interactions, both phases as well as the critical point are compressible and this implies an exact dynamical critical exponent $z = 2$. (With Coulomb interactions, $z$ is modified to $1$). Moreover, by the Harris-Chayes bound~\cite{Harris1974_bound,Chayes1986_bound}, the correlation length exponent $\bar \nu$ satisfies the inequality $\bar \nu \geq 1$ in two dimensions. As a result, $\bar \nu z \geq 2$ and the width of the resistance peak shrinks as a slow power law of $T$.

For $\mu_{c1} < \mu < \mu_{c2}$, the AVG phase has a non-vanishing resistance at $T > 0$ that gradually decreases to zero as $\mu$ approaches the second transition at $\mu_{c2}$. If the localized anyon density is small at $\mu = \mu_{c1}$, the AVG phase occupies a very narrow region on the $\mu$-axis and an apparent direct transition between the FQAH anyon glass and the ordinary SC would occur at moderate temperature. In that case, the scaling of $\rho_s(\mu)$ follows an approximate scaling form
\begin{equation}
    \rho_s \sim \xi^{-z} \sim |\mu - \mu_{c1}|^{\bar \nu z} \,. 
\end{equation}This phenomenology is summarized in Fig.~\ref{fig:FQAH_SC_smooth_rhos} and Fig.~\ref{fig:FQAH_SC_smooth_rho}. 

The story is drastically different when the microscopic disorder potentials $W_{IJ}(\bs{r})$ and $W_b(\bs{r})$ have significant variations at the lattice scale. In this case, all components of $W_{IJ}$ must be included and the emergent $SU(3)$ symmetry in the clean Lagrangian is completely broken. As $\delta$ increases, the evolution from $\sigma_d^{xy} = 0$ to $\sigma_d^{xy} = -3$ passes through three separate phase transitions, as $\sigma_d^{xy}$ can only jump by $\pm 1$ at each critical point. When $\sigma_d^{xy} = -1$, the effective Lagrangian reduces to 
\begin{equation}
    L[\sigma_d^{xy} = -1] = \frac{2}{4\pi} bd b \,, 
\end{equation}
which describes a localized insulator that coexists with a neutral $U(1)_{-2}$ topological order. This phase is indistinguishable from a trivial localized insulator in electric transport, but has a single neutral chiral edge mode with $\kappa_{xy} = - 2 \kappa_Q$ that can potentially be detected in thermal Hall measurements.\footnote{This state can also be obtained by doping the CDW* state with neutral $U(1)_{-2}$ topological order at $\nu = 2/3$ which is accessible from the Jain state through a bandwidth-tuned transition~\cite{Song2023_QPT_FQAH}.}  
When $\sigma_d^{xy} = -2$, the effective Lagrangian becomes
\begin{equation}
    L[\sigma_d^{xy} = -2] = \frac{1}{4\pi} bd b + \frac{1}{2\pi} b d A - \frac{1}{4\pi} A d A - 2 \mathrm{CS}_g \,. 
\end{equation}
Integrating out $b$ simplifies this Lagrangian further to
\begin{equation}
    L[\sigma_d^{xy} = -2] = - \frac{2}{4\pi} A d A - 4 \mathrm{CS}_g \,, 
\end{equation}
which describes an IQH state with Hall conductance $\sigma^{xy} = -2$.  
Putting all of these results together, we arrive at the phase diagram in Fig.~\ref{fig:FQAH_SC_rough}, where each critical point is associated with a universal $\mathcal{O}(h/e^2)$ resistivity tensor. At $T>0$, the width of the transition region scales as $T^{1/(\bar \nu z)}$, while the critical resistivity is independent of $T$. Importantly, although the three critical points involve the same jump in $\sigma_d^{xy}$ at the mean field level, they differ in the structure of gauge interactions and could have distinct critical exponents. If we decrease the disorder strength at fixed $T$, the range of $\delta$ over which the three transitions occur shrinks and the peaks eventually merge into a smoother evolution of $\rho^{xx}$ and $\rho^{yx}$. 

\paragraph*{Doping the $a_{1/3}$ anyon;}
We now turn to the second possibility, in which the doped anyon is $a_{1/3}$ and the $f_1$ band is closest to the chemical potential. In the clean limit, we again introduce three species of $d_I$ that create excitations of $f_1$ near its band minima at $\bs{K}_I$ such that $d_I$ transforms projectively under $T_x, T_y$ as in \eqref{eq:d_Iproj}. The low energy effective Lagrangian then takes the form
\begin{equation}
    \begin{aligned}
    &L_{\rm clean, a_{1/3}} \\
    &= \sum_{I=1}^3 d_I^{\dagger} \left[i \partial_t + \mu + a_t - \frac{(i\nabla + \bs{a})^2}{2m}\right] d_I + L_{\rm int}[d_I] \\
    &\hspace{0.4cm} - \frac{1}{4\pi} a da + \frac{2}{4\pi} bdb + \frac{1}{4\pi} A d A - \frac{1}{2\pi} b d (A + a) \,,
    \end{aligned}
\end{equation}
where $m$ now denotes the effective mass of $f_1$ near the band minima. When $\mu < 0$, $d_I$ is gapped and we recover the Jain state. When $\mu > 0$, the equations of motion for $a_t$ and $b_t$ enforce the constraint $\sum_I \rho_I = \frac{3}{2} \frac{1}{2\pi} \ev{\nabla \times \bs{a}}$. At the mean-field level, each $d_I$ feels an effective magnetic field $\nabla \times \bs{a}$ and is at Landau level filling 1/2. Therefore, the three species of $d_I$ form three pockets of composite Fermi liquids coupled to a common fluctuating gauge field $a$. The analysis in Ref.~\cite{Shi2024_doping} shows that gauge fluctuations induce interpocket pairing between two out of the three pockets and confine all the fluctuating gauge fields. The resulting ground state is a charge-ordered Fermi liquid coexisting with a background IQH state
\begin{equation}\label{eq:L_FLclean}
    L_{\rm FL} = L_{\rm FL}[c, A] + \frac{1}{4\pi} A d A + 2 \mathrm{CS}_g \,. 
\end{equation}
The clean critical theory is identical to the clean FQAH-SC transition at the mean field level and the argument for decoupling between gauge and matter sectors still holds at $T = 0$. At $T > 0$, this transition differs from the FQAH-SC transition since the relevant gauge fields have different Chern-Simons terms. A quantitative understanding of these finite-temperature properties is an interesting target for future studies.

The story has an interesting twist when we include quenched disorder. With long-wavelength disorder, the random potential Lagrangian still takes the $SU(3)$ symmetric form in \eqref{eq:Ldis_smooth}. At low doping, the resulting state is an anyon glass (with quantized $\sigma^{xy} = 2e^2/3h$) made of $a_{1/3}$ instead of $a_{2/3}$. Unlike in the FQAH-SC case, the mean-field transition is not an IQH plateau transition of $d_I$. Instead, it describes the  creation of three composite fermion pockets out of a localized insulator, about which very little is understood. Moreover, on the other side of the phase transition, the Fermi liquid with period-3 charge order predicted by the clean theory will not be stable to localization effects at low doping density. In the  resulting state, the $L_{\rm FL}[c,A]$ part of \eqref{eq:L_FLclean} becomes a Fermi glass and the theory describes an IQH state with quantized $\sigma^{xx} = 0$ and $\sigma^{xy} = e^2/h$. Further in the approximation that the smooth disorder potential does not couple linearly to the CDW order parameter, the charge order will survive in this IQH state.

With short-wavelength disorder, the random potential breaks the $SU(3)$ symmetry. At finite doping the charge order of the doped Fermi liquid will be unstable to disorder. As before  at low doping densities the Fermi liquid will again be unstable to localization. The integer quantum Hall effect of the clean CDW metal will survive. Let us now consider the evolution from the FQAH to this IQH phase. The individual filling factors of $d_I$ are no longer meaningful and we can regard $d_I$ as a three-component fermion at total Landau level filling $3/2$. At low doping, $d_I$ forms a localized insulator and we again have an anyon glass of $a_{1/3}$. This of course is part of the FQAH plateau. Upon increasing the doping, it is natural to encounter a plateau transition in which the total Hall conductance of the $d_I$ sector jumps by $1$. When this occurs, the effective Lagrangian becomes
\begin{equation}
    \begin{aligned}
    L &= \frac{2}{4\pi} bdb + \frac{1}{4\pi} A d A - \frac{1}{2\pi} b d (A+a) + 2 \mathrm{CS}_g \,.
    \end{aligned}
\end{equation}
Integrating out $a$ sets $b = 0$ and gives
\begin{equation}
    L[\sigma_d^{xy} = 1] = \frac{1}{4\pi} A d A + 2 \mathrm{CS}_g \,, 
\end{equation}
which just describes an IQH state coexisting with a localized Fermi glass. As argued above, this is exactly the fate of the clean CDW metal (+IQH) at low charge densities in the presence of short wavelength disorder.

Therefore, the most natural state that arises from doping in $a_{1/3}$ is an IQH state with quantized $\sigma^{xy} = e^2/h$, although the nature of the transition from the Jain state to this IQH state could depend on the type of quenched disorder. A schematic transport plot is shown in Fig.~\ref{fig:FQAH_RIQAH_rho}.

\paragraph*{Bandwidth-tuned FQAH-SC transition:} We now briefly consider the possibility that the FQAH state can be be tuned to a SC at fixed lattice filling $\nu = 2/3$ by varying the bandwidth (e.g. with a displacement field) in the clean limit.  To that end, we begin with the parton description of the FQAH state in terms of $f_i$ for $i = {1,2,3}$ each filling a Chern band with Chern numbers $C_i$ such that $C_1 = -2, C_2 = C_3 = 1$. We consider a phase transition at fixed filling where the Chern number of $f_3$ changes away from $1$ (through band inversion) while the $f_1, f_2$ Chern numbers stay fixed.   
As we discussed, the projective action of translation symmetry implies that there are three inequivalent points in the Brillouin zone that are degenerate in energy. Consequently, the band touching happens simultaneously at these three points, each described by a massless Dirac fermion. The $f_3$ Chern number can therefore only jump by multiples of $3$ so long as the lattice translation symmetry is preserved. 

\begin{figure}
    \centering
    \includegraphics[width=\linewidth]{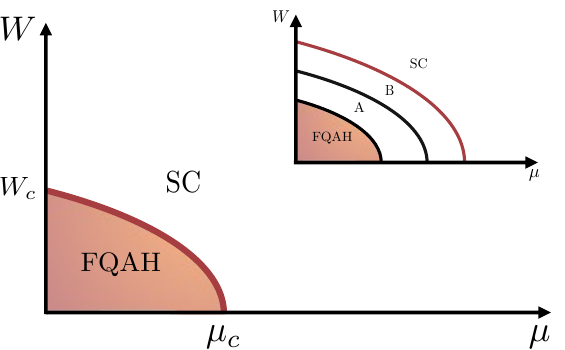}
    \caption{Schematic phase diagram showing the $2/3$ Jain FQAH and chiral topological SC in the bandwidth-chemical potential ($W-\mu$) plane. In a clean system, the phase transition at fixed lattice filling is in a different universality class described by the QED$_3$-Chern Simons theory of \eqref{eq:L_dirac}. Inset: Dirty limit with short-wavelength disorder. The transition goes through two intermediate phases; phase (A) is an insulator with a neutral topological order and (B) is an integer quantum Hall insulator with $\sigma^{xy} = - 2 e^2/h$.}
    \label{fig:fqahscDmu}
\end{figure}

Consider now a phase where the Chern number of $f_3$ has changed from its original value $C_3 = 1$ in the FQAH phase to $C_3 = -2$. The low energy effective theory of the resulting phase is 
\begin{eqnarray} 
    L & = & \frac{2}{4\pi } bdb - \frac{2}{4\pi} (A-b) d(A - b) -4  \mathrm{CS}_g \\
    & = & \frac{2}{2\pi} b d A - \frac{2}{4\pi} AdA - 4 \mathrm{CS}_g \,.
\end{eqnarray} 
Integrating out $b$, we see that we get a charge-$2$ superconductor with a chiral central charge $c_- = -2$, which is smoothly connected to the one obtained by doping the FQAH state with the $a_{2/3}$ anyon~\eqref{eq:L_SCclean}. At the phase transition, the critical Lagrangian takes the form
\begin{equation}\label{eq:L_dirac}
    \begin{aligned}
    L_{\rm Dirac} &= \sum_{I=1}^3 d_I^{\dagger} \left(i\slashed{\partial} + \slashed{A} - \slashed{b}\right)  d_I \\
    &\hspace{0.2cm}   + \frac{3}{4\pi} b db - \frac{1}{2\pi} b d A + \frac{1}{4\pi} A d A + 2 \mathrm{CS}_g \,.
    \end{aligned}
\end{equation}
This is a QED$_3$ theory with $N_f = 3$ massless Dirac fermions (representing the 3 band touching points of $f_3$) coupled to an emergent gauge field $A-b$. As a relativistic theory, it has $z = 1$ and power-law correlations in the CDW order parameters $d^\dagger_I T_{IJ} d_J$ (with $T_{IJ}$ an $SU(3)$ matrix). Its properties are similar to an analogous transition between the bosonic $\nu = 1/2$ Laughlin and superfluid states, which has $N_f = 2$ and distinct Chern-Simons terms. In Appendix~\ref{app:bosonic_jain}, we show that a useful perspective on this transition as well as the doping-induced FQAH-SC transition can be obtained by relating them to transitions out of the bosonic Jain state at $\nu = 2/3$.

A schematic phase diagram in the full bandwidth-chemical potential plane is shown in Fig.~\ref{fig:fqahscDmu}. 

\paragraph*{Application to doped \textit{t}MoTe$_2$:}

A recent experimental study~\cite{Xu2025_SCdopeTMD} of a high quality hole-doped \textit{t}MoTe$_2$ device explored the phase diagram in the vicinity of the prominent $\nu = 2/3$ FQAH state. In a small range of displacement field near zero, a SC state is found at fillings $0.71 \lesssim \nu \lesssim 0.76$. This SC is separated from the FQAH state by a region where the longitudinal resistance $\rho^{xx}$ peaks with a peak height of order $h/e^2$ (in the range $10-15 ~k\Omega$). On the other side with $\nu$ just smaller than $2/3$, an integer quantum Hall response is found (dubbed the Rentrant Integer Quantum Anomalous Hall (RIQAH) state) in $\rho^{xy}$ and $\rho^{xx}$. This too is separated from the FQAH state by a peak in $\rho^{xx}$. Caution is however needed in interpreting the RIQAH response: a plot of the Streda slope in this state yields a $\sigma^{xy}$ closer to $2/3$ than to $1$. This discrepancy may indicate issues~\cite{Kane1994_random_edge,Kane1995_edge_eq,Kane1995_contact,Chamon1997_contact,Young_private} with edge equilibration in the multi-component edge expected for the $2/3$ FQAH state which could lead to a spurious value of the measured $\rho^{xy}$.  An RIQAH response is also seen at higher displacement field for $\nu \gtrsim 2/3$ (with the same concerns as above with the Streda slope). 

It is important for future experiments to resolve this discrepancy between transport and the Streda slopes, and to have further confirmation of the superconducting state. We have argued that it is reasonable {\it theoretically}  to expect both superconductivity and an RIQAH state in proximity to the $2/3$ FQAH plateau. Let us therefore examine the existing experimental data in light of our theory.

There are several preliminary questions that need to be settled before interpreting these observations in terms of the theory developed in this paper. For the observed superconductor, it is important to know if the mobile charge carrier density is ``small" (corresponding to filling $2/3 - \nu$, the deviation from $2/3$) or if it is ``large" (corresponding to filling $1 - \nu$). If the former, then a description as a doped FQAH state - which will naturally involve doped anyons - is the right framework. If the latter, then it is appropriate to think of the SC as developing out of a normal state that is separated from the FQAH state by a strong first order transition, and the theory described in this paper is not directly pertinent. The existing data reveals a sign change of the Hall effect at high fields $B \gtrsim 2~ T$ near $\nu \approx 0.75$ at $D = 0$ below which SC first develops. This could be consistent with a `large' to `small' carrier density change. More detailed studies - as sketched in Appendix \ref{app:anyongas} -  are needed to determine whether the carrier density truly decreases towards zero as $\nu$ approaches $2/3$.

Assuming that the doped FQAH framework is applicable,  the theory developed so far provides a possible explanation of these observations. At displacement field $D = 0$ and $\nu = 2/3$, we propose that the anyon energetics of the Jain state has a particle-hole asymmetry, such that doping towards $\nu > 2/3$ induces a fluid of $a_{2/3}$, while doping towards $\nu < 2/3$ induces a fluid of $a_{1/3}$. This translates to a parton band structure where the conduction (valence) band closest to the chemical potential is associated with $f_3$ ($f_1$). With long-wavelength disorder\footnote{In high quality \textit{t}MoTe$_2$ devices, the density of charged point defects (which generate short-wavelength disorder) is about a factor of 100 smaller than the electron density~\cite{Heonjoon_private}. It is thus plausible that other sources of disorder (such as a random twist angle), which are only modulated at long wavelengths, play a dominant role.}, there is a narrow plateau with quantized $\sigma^{xy} = 2 e^2/3h$ for $\nu \in [\nu_1, \nu_2]$ with $\nu_1 < 2/3 < \nu_2$. Doping beyond the plateau, our theory predicts a topological superconductor at $\nu > \nu_2$ and an RIQAH state at $\nu < \nu_1$. While all three phases have $\rho^{xx} = 0$, the transitions between them feature $\mathcal{O}(h/e^2)$ peaks in $\rho^{xx}$, broadly consistent with the experimental observations in Ref.~\cite{Xu2025_SCdopeTMD}. In the $T \rightarrow 0$ limit, the height of the $\rho^{xx}$ peak at either transition is expected to saturate to a non-zero value while the width should decrease to zero as we described earlier. It would be interesting for future experiments to clarify these features of the resistance peaks. 

Near the initial onset of superconductivity at $T = 0$ out of the anyon glass, we argued that the superconducting state is an Anomalous Vortex Glass with a random sprinkling of $-2\pi$ and $+4\pi$ vortices with net zero vorticity. At $T > 0$, the AVG has a non-zero linear resistance which goes to zero only gradually as the temperature is decreased toward zero. This feature and the expected non-linear response at small current bias are also seen in the experiments reported in Ref. \cite{Xu2025_SCdopeTMD}. Further studies are needed to establish whether these features are due to the AVG or have other mundane explanations.

An important future direction is a microscopic calculation of anyon energetics within models of \textit{t}MoTe$_2$. It will be interesting to see if ``anyon bunching'' of the kind reported in very recent calculations~\cite{Xu2025_anyon_bunching,gattu2025molecularanyonsfractionalquantum} of Landau level FQH states happens in FQAH states found in models of \textit{t}MoTe$_2$ (see Ref. \cite{Miguel2025_anyon_dispersion} and the note added).  However, even without numerics, existing data already provide some qualitative guidance. Given the sequence of Jain states observed near $\nu = 1/2$, it is natural to expect that the $\nu < 2/3$ side fits into the standard ``composite fermion" framework~\cite{Jain1989_CFframework}. Since $f_1$ is the zero-field analogue of the composite fermion (see Ref.~\cite{Goldman2023_ACFL,Dong2023_ACFL,Shi2024_doping}), it is reasonable to associate the nearest valence band with $f_1$. On the other hand, for $\nu > 2/3$, there is no a priori reason for the composite fermion framework to apply. Thus, our proposal of $f_3$ being the nearest conduction band becomes more plausible.  

Tuning away from $D = 0$ modifies the anyon dispersion at $\nu = 2/3$. If there exists a critical $D_c$ such that the nearest conduction band switches from $f_3$ to $f_1$ for $D>D_c$, then doping induces a fluid of $a_{1/3}$ anyons for both $\nu > 2/3$ and $\nu < 2/3$. This switching effect provides a potential explanation for the pair of RIQAH states observed in a range of displacement fields $15 \lesssim D \lesssim 30$ nV/nm on both sides of the FQAH plateau~\cite{Xu2025_SCdopeTMD}.

It will also be important to have an experimental determination of the charge of the delocalizing anyon. This may be easier in the FQAH/anyon glass plase prior to the delocalization transition. Standard measurements~\cite{saminadayar1997observation,de1998direct,martin2004localization} of the anyon charge successfully used in ordinary FQH systems are of course worth exploring in the FQAH context. An indirect route, specific to the lattice setting, is to look for local charge density modulations that are expected~\cite{song2024density} to appear around each localized anyon (with short wavelength disorder) in a halo region at low dopant density. This is due to the close competition between FQAH and Charge Density Wave (CDW)  phases in a clean system. Around each anyon, the FQAH is locally suppressed enabling the CDW to rear its head. These modulations may be measurable in Scanning Tunneling Microscopy experiments and their density can potentially yield the anyon charge~\cite{song2024density}. 

Finally our discussion of the transmutation of anyons of the FQAH into vortices of the SC could help test the basic validity of the anyonic mechanism of superconductivity. Following similar discussions in the context of the cuprates~\cite{senthil2001fractionalization,senthil2001bfractionalization,bonn2001limit}, vortex memory effects are expected when cycling an annulus between the superconductor and the FQAH state. Specifically $\pm h/2e$ vortices trapped in a hole inside the device would first disappear when the doping is tuned back into the insulator but would leave behind a threaded anyon. Upon cycling back into the SC, the anyon would nucleate the same vortex.

\textit{Note added}—We are coordinating submission with Ref.~\cite{Nosov2025_dope}, where closely related results were obtained. Since the original posting of this paper on the arXiv, Refs.~\cite{zhang2025holon,pichler2025microscopicmechanismanyonsuperconductivity} appeared on related topics, and Ref.~\cite{Miguel2025_anyon_dispersion} on some pertinent microscopic calculations.

\paragraph*{Acknowledgements:} We thank Matthew Fisher, Tonghang Han, Eslam Khalaf, Tingxin Li, Pavel Nosov, Heonjoon Park, Nicolas Regnault, and Ashvin Vishwanath for discussions. ZDS and TS are supported by NSF grant DMR-2206305.

\bibliography{theory_tMoTe2} 

\onecolumngrid
\appendix

\newpage 

\section{Connection between the transitions out of bosonic/fermionic Jain states}\label{app:bosonic_jain}

In the main text, we analyzed various possible paths from the fermionic Jain state at $\nu = 2/3$ to a chiral topological superconductor with $c_- = -2$ by doping in the $a_{2/3}$ anyon. In this section, we provide an alternative perspective on these transitions by relating the fermionic Jain state to a bosonic Jain state of charge-$2$ Cooper pairs at the same filling stacked with a fermionic integer quantum Hall state. The possibility of such a mapping is due to the fact that powers of $a_{2/3}$ cannot generate the physical electron. As a result, the low energy Hilbert space is bosonic and various transitions admit a purely bosonic description. In this Appendix, we will develop this description which will give an interesting alternate theoretical viewpoint on the FQAH-SC evolution described in the main text. See also Ref. ~\cite{Divic2024_anyonSC} for a recent study of superconductivity and topological quantum criticality in an electronic Hofstadter-Hubbard model that can be given a similar interpretation. 

The fermionic $2/3$ Jain state has Hall conductivity $\sigma_{xy}^f = 2e^2/3h$ and a chiral central charge $c_-^f = 0$.  The  3 distinct anyons generated by $a_{2/3}$ are  $(1, a_{2/3}, a_{2/3}^2)$ which have statistics $(0, 2\pi/3, 2\pi/3)~\mod 2\pi$ respectively. The corresponding electric charges are $0, 2/3, 4/3$. These anyons form a complete {\it bosonic} topological order by themselves (which we denote ${\cal V}^{3,2}$). The full anyon content of the fermionic $2/3$ Jain state can be described as the tensor product ${\cal V}^{3,2} \times (1,c)$ where $c$ is the physical electron. 

Consider now a different theory, namely, the bosonic $2/3$ Jain state of charge-$2$ Cooper pairs\footnote{In a Landau level setting this will be described as the Halperin $(2,2,1)$ state. The corresponding $2 \times 2$ $K$-matrix has $K_{11}=K_{22} = 2, K_{12} = K_{21} = 1$ and a charge vector $(2,2)$.}. This has electrical Hall conductivity $\sigma^b_{xy} = 8e^2/3h$, and a chiral central charge $c_{-}^b = 2$. Following standard arguments, this will have a vison $v$ (the anyon nucleated by $2\pi$ flux threading) with statistics $2\pi/3$ and electric charge $8/3$. The full topological order has 3 anyons $(1,v,v^2)$ with statistics $(0, 2\pi/3, 2\pi/3)\mod 2\pi$ respectively. The corresponding electric charges are $(0, 8/3, 16/3) \simeq (0, 2/3, 4/3)\mod 2$. The modding by $2$ corresponds to binding Cooper pairs to get  topologically equivalent particles. 

Thus the anyon data of the bosonic Jain state exactly match that of ${\cal V}^{3,2}$ but the electrical conductivity and chiral central charge differ from the fermionic $2/3$ Jain state: $\sigma^f_{xy} = \sigma^b_{xy} - 2e^2/h$, $c_-^f = c_-^b -2$. But this difference is precisely that of a $\sigma^{xy} = -2$ fermionic integer quantum Hall state. Thus we can view the fermionic $2/3$ Jain state as a bosonic $2/3$ Jain state of charge-$2$ Cooper pairs stacked with a $\nu = -2$ fermionic integer quantum Hall state. For other similar identifications, see the recent discussion in Ref. \cite{Cheng2025_orderingqh}. 

We can also obtain the same conclusion through an explicit field theoretic construction. Let us begin by considering a microscopic system of charge-2 bosons $\Phi$ at lattice filling $2/3$. To describe the bosonic Jain state, we consider a parton decomposition $\Phi = f_1 f_2$ and introduce an emergent $U(1)$ gauge field $a$ to remove the gauge redundancy. The parton Lagrangian takes the form
\begin{eqnarray}
    L[\Phi, 2A] = L[f_1, 2A - a] + L[f_2, a] \,. 
\end{eqnarray}
We interpret the charge-2 boson as the bound state of two physical fermions, each carrying charge 1. In this case, $A, a$ are $\mathrm{spin}_{\mathbb{C}}$ connections and $2A$ is an ordinary $U(1)$ gauge field. At the mean-field level, we choose the flux of $a$ to be proportional to $2\pi/3$ so that $f_1, f_2$ can form Chern insulators with Chern numbers $C_1, C_2$. To construct the bosonic Jain state, we take $C_1 = 2, C_2 = 1$. Integrating out the gapped $f_1, f_2$ then produces an effective Lagrangian
\begin{eqnarray}
    L_{\rm bosonic, 2/3} = \frac{2}{4\pi} (2A - a) d (2A-a) + \frac{1}{4\pi} a d a + 6 \mathrm{CS}_g \,. 
\end{eqnarray}
Now we make a change of variables $a = b + A$ so that $b$ is an ordinary $U(1)$ gauge field. The resulting Lagrangian simplifies to
\begin{eqnarray}
    L_{\rm bosonic, 2/3} = \frac{2}{4\pi} (A - b) d (A-b) + \frac{1}{4\pi} (b + A) d (b + A) + 6 \mathrm{CS}_g = \frac{3}{4\pi} bdb - \frac{1}{2\pi} b d A + \frac{3}{4\pi} A d A + 6 \mathrm{CS}_g \,. 
\end{eqnarray}
The difference between this Lagrangian and the Lagrangian in \eqref{eq:Jain_TQFT} for the fermionic Jain state is precisely the response theory for a $\nu = -2$ fermionic IQH state. We therefore conclude that
\begin{eqnarray}\label{eq:relation_bJain_fJain}
    L_{\rm fermionic, 2/3} = L_{\rm bosonic, 2/3} - \frac{2}{4\pi} A d A - 4 \mathrm{CS}_g \,. 
\end{eqnarray}

The above matches the topological order and the charge assignments of the fermionic $2/3$ Jain state and the bosonic one stacked with a $\nu = -2$ fermion IQHE. The projective translation symmetry action on the anyons will also be the same. This is because at lattice filling $2/3$, there will be a background anyon in each unit cell which is the charge $2/3$ anyon. The translation symmetry action on any  anyon is completely determined by its braiding phase around the background anyon, which in turn is determined by its fractional charge. Since we have already matched the charges, the translation action will also match. 

\begin{figure}
    \centering
    \includegraphics[width=0.8\linewidth]{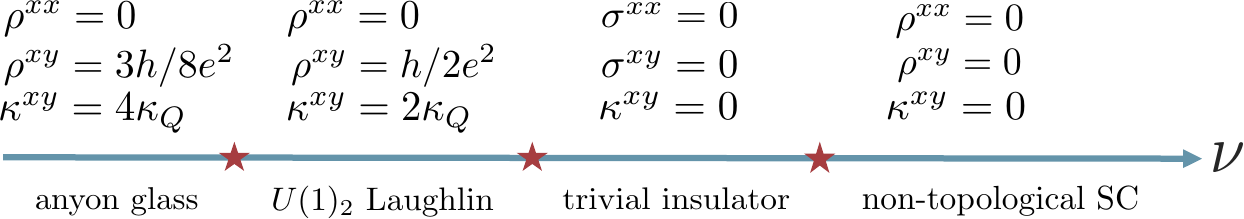}
    \caption{Phase diagram of the doped bosonic Jain state in the dirty limit with short-wavelength disorder. Note that the four phases here can be obtained from the four phases in Fig.~\ref{fig:FQAH_SC_rough} by stacking with a $\nu = -2$ fermionic IQH state.}
    \label{fig:bosonicJain}
\end{figure}

Within the bosonic theory, doping the $a_{2/3}$ anyon corresponds to doping into the $f_1$ band. When the disorder potential has large short-wavelength modulations, the evolution from the bosonic Jain state to the trivial superconductor passes through four stages, in which the Hall conductance of $f_1$ is $\sigma_{f_1}^{xy} = 2, 1, 0, -1$ respectively. When $\sigma_{f_1}^{xy} = 2$, we recover the bosonic Jain state at $\nu = 2/3$. When $\sigma_{f_1}^{xy} = 1$, we obtain 
\begin{eqnarray}
    L = \frac{1}{4\pi} (2A - a) d (2A - a) + \frac{1}{4\pi} a da + 4 \mathrm{CS}_g \,.
\end{eqnarray}
We recognize this Lagrangian as the standard parton description of the bosonic $U(1)_2$ state with charge-2 bosons. When $\sigma_{f_1}^{xy} = 0$, the Lagrangian reduces to 
\begin{eqnarray}
    L = \frac{1}{4\pi} a da + 2 \mathrm{CS}_g \,, 
\end{eqnarray}
which describes a trivial insulator. Finally, when $\sigma_{f_1}^{xy} = -1$, we can again make a change of variables $a = b + A$ to get an effective theory
\begin{eqnarray}
    L_{\rm bosonic, 2/3}[A] = - \frac{1}{4\pi} (A - b) d (A - b) + \frac{1}{4\pi} (b+A) d (b + A) = \frac{2}{2\pi} b d A \,. 
\end{eqnarray}
Therefore, the bosonic sector describes a trivial charge-2 superconductor. This four-stage evolution is summarized in Fig.~\ref{fig:bosonicJain}. In the clean limit (or with smooth disorder), two of these intermediate phases with $\sigma_{f_1}^{xy} = 1$ and $\sigma_{f_1}^{xy} = 0$ disappear and we are left with a direct transition between the bosonic Jain state and the trivial charge-2 superconductor. 

As a sanity check, we can verify that the four phases (see Fig.~\ref{fig:FQAH_SC_rough}) in the doped fermionic Jain state can be directly obtained from the four phases (see Fig.~\ref{fig:bosonicJain}) in the doped bosonic Jain state by stacking with a $\nu = -2$ integer quantum Hall state of electrons. This is of course consistent with the relation in \eqref{eq:relation_bJain_fJain}. 

Since the background $\nu = -2$ integer quantum Hall response does not influence the dynamics of the system, the transitions between neighboring phases in Fig.~\ref{fig:FQAH_SC_rough} lie in the same universality class as the corresponding transitions in Fig.~\ref{fig:bosonicJain}. This observation provides additional insight into the nature of these transitions. For example, the transition between the $U(1)_{-2}$ neutral topological order and the $\nu = -2$ IQH state in the fermionic formulation is mapped to a transition between the $U(1)_2$ bosonic Laughlin state and a trivial insulator. The latter transition admits a simple Chern-Simons description in the clean limit, which can be studied numerically in the presence of disorder. Similarly, the transition between the $\nu = -2$ IQH state and the chiral topological SC in the fermionic formulation is mapped to a standard superfluid-insulator transition in the bosonic formulation, about which a lot of progress has been made since the pioneering works of Refs.~\cite{fisher1989boson,fisher1990presence}. 

Finally, in a clean system,  after removing the stacked IQH states, the bandwidth tuned fermionic $2/3$ Jain-chiral topological SC gets exactly related to the bosonic $2/3$ Jain to the trivial SC transition. This latter transition is part of the family of bosonic $p/(p+1)$ Jain - trivial SC transitions which are described by $p+1$ massless Dirac fermionc coupled to a $U(1)$ gauge field with a  Chern-Simons action: 
\begin{equation}\label{eq:L_diracB}
    \begin{aligned}
    L_{\rm DiracB} &= \sum_{I=1}^{p+1} d_I^{\dagger} \left(i\slashed{\partial} + \slashed{a} \right)  d_I \\
    &\hspace{0.2cm}   + \frac{p+1}{4\pi} a da - \frac{2}{2\pi} a d A + \frac{4}{4\pi} A d A + 2(p+1) \mathrm{CS}_g \,.
    \end{aligned}
\end{equation}
(We have assumed that the physical bosons have charge-$2$). Setting $p = 2$, and letting $a = A-b$, we see that we recover Eqn. ~\ref{eq:L_dirac} after removing the $\nu = 2$ fermionic IQH response. 

\section{Anyon-induced vortices in the superconductor}\label{app:AVG}

In this Appendix, we study the relationship between anyons in the parent FQAH state and vortices in the superconducting state. This will enable us to understand how the expected $h/2e$ flux quantization of the charge-$2$ condensate emerges in this anyon-induced superconductor. It will also enable us to discusss the fate of localized anyons as the anyon glass transitions into the superconductor upon increasing the doping. 

To set the stage, we consider an annular geometry with a single anyon threading the inner hole in the FQAH phase. What happens when we move into the superconducting state? To address this, it is useful conceptually to take the viewpoint on the doped SC developed in Ref. \cite{Shi2025_doping_nonabelian}. We start with the $2/3$ fermionic Jain state, and stack a neutral bosonic $-2/3$ Jain state on top. The latter has an anyon (which we denote $v_b$) with statistics $-2\pi/3$ which generates the full topological order. The anyon $a_{2/3} v_b$ is then a charge $2/3$ boson and condensing it gives a SC state. Furthermore, it can be checked that the condensation process removes all other anyons so that we have an ordinary SC. Finally as the bosonic $-2/3$ Jain state has $c_- = -2$ while the fermionic $2/3$ Jain state has $c_- = 0$, the SC will have $c_- = -2$ in agreement with the field theory analysis. 

Now consider the annulus with a single $a_{2/3}$ threading the inner hole. The condensing boson $\Phi \equiv a_{2/3}v_b$ winds by $4\pi/3$ around $a_{2/3}$. Thus if we let $\Phi = e^{i\theta}$ with $\theta$ classical, then $\theta$ is not single-valued as we go around any loop ${\cal C}$ that encloses the inner hole, and we have 
\begin{equation} 
\int_{\cal C} d \vec l \cdot \vec \nabla \theta = \frac{4\pi}{3} + 2n\pi 
\end{equation} 
(with $n$ an integer). The Cooper pair field $\Phi_c = \Phi^3$ is however single-valued. Defining the Cooper pair phase through $\Phi_c = e^{i\theta_c} = e^{3\theta}$, we get 
\begin{equation} 
\int_{\cal C} d \vec l \cdot \vec \nabla \theta_c  = 4\pi + 6n\pi 
\end{equation} 
Thus a superconducting vortex is nucleated\footnote{See Refs. \cite{senthil2000z,senthil2001fractionalization} for a similar discussion for superconductivity emerging in doped $\mathbb{Z}_2$ spin liquids.}   with vorticity $-2\pi, 4\pi,\cdots$. Which of these vortices is nucleated will be determined by energetic considerations. 

If instead we had an $a_{1/3}$ anyon threading the inner hole in the FQAH, then the condensing boson in the proximate superconductor winds by $8\pi/3 \equiv 2\pi/3 \mod 2\pi$. Consequently a vortex of strength $2\pi \mod 6\pi$ in the physical Cooper pair phase is nucleated.  

We can also give a field-theoretic discussion of this phenomenon. In the FQAH phase, the $a_{2/3}$ anyon is nucleated by the $f_3$ fermionic parton which couples to $A - b$. Upon transitioning to the SC, with an effective action described by \eqref{eq:L_SCclean}, a source of $A - b$ becomes a vortex by the usual particle-vortex duality, and it is easy to see that it is a $-2\pi$ vortex of the pair order parameter. To see the periodicity of $6\pi$, we note that in the parent FQAH a field that carries charge 3 under $b$ is local, and has physical charge 1. This quasiparticle is identified with the microscopic electron. Thus we can describe the same $a_{2/3}$ anyon as a bosonic source $\phi$ of $b$ with charge 2 rather than as a fermionic source for $A-b$ with unit charge, and this changes the vorticity by $6\pi$.

Now consider the superconducting state emerging out of the anyon glass with a finite density $n_{loc}$ of localized $a_{2/3}$ anyons. We first give a physical discussion. 
Around a  single localized anyon, if, say, a $-2\pi$ vortex is nucleated, it has to pay the logarithmic superflow energy cost which diverges in the thermodynamic limit and overwhelms the energy gain by keeping the vortex localized in the random potential. However, with a nonzero density of anyons present, we can sprinkle $-2\pi$ vortices and $4\pi$ vortices together so that the net vorticity is zero. In this case, the superflow energy per localized anyon would be finite and of order $\rho_s \ln{1/n_{loc}}$ (in units where the lattice spacing is 1) which for small $\rho_s$ can be smaller than the $\mathcal{O}(1)$ random potential energy gained by keeping the vortex localized. Thus for small $\rho_s$, {\em i.e.} close to the onset of superconductivity out of the anyon glass, we expect that the $T= 0$ superconducting state is a vortex glass~\cite{fisher1989vortex,fisher1991thermal,fisher1991vortex}. Unlike the usual vortex glass, this state occurs at zero magnetic field, and we will therefore call it an ''Anomalous Vortex Glass'' (AVG). 

Let us discuss this state more formally by considering \eqref{eq:L_SCclean} in the presence of static localized anyons. We describe the localized $a_{2/3}$ anyons through a random profile of the gauge charge density that couples to the $b$ field. Following the discussion in the previous paragraph, we should include localized sources of both $A-b$ and $2b$ as they correspond to the same $a_{2/3}$ anyon but become inequivalent in the superconducting state. To determine the optimal configuration of the superconducting phase, it is sufficient to consider a classical energy (rather than action) density functional involving just the time-component $b_0$ of the $b$ gauge field 
\begin{equation}
 \tilde{E}_{\rm SC} =  \frac{1}{8\pi^2 \rho_s} \left(\nabla  b_0 \right)^2  - j_0^f b_0 + 2j_0^\phi b_0   \,
 \end{equation} 
 Here $j_0^f + j_0^b = \sum_n \psi_n(\vec r - \vec r_n)$ is the density profile of the localized anyons, and $\psi_n$ is a localized function that represents the density distribution around each random site $n$. We have included a Maxwell term for the $b_0$ gauge field that penalizes `electric field' fluctuations with a coupling constant  $\rho_s$. By usual duality arguments, this model can be recast as a classical $XY$ model for the superconducting phase $\theta_c$ coupled to a random magnetic field. Specifically, we write $j_0^{b,f} = \frac{\nabla \times \bs{a}^{b,f}}{2\pi}$. Then the dual version of the above energy functional becomes 
 \begin{equation} 
\tilde{E}_{\rm SC} = \frac{\rho_s}{2} \left(\nabla \theta_c - \bs{a}^f + 2 \bs{a}^b \right)^2  
 \end{equation} 
Here $\theta_c \sim \theta_c + 2\pi$ so that vortices are included. The sum of the fluxes of the random static gauge fields $\int \frac{1}{2\pi} \left(\nabla \times (\bs{a}^f + \bs{a}^{b})\right)$ is given by the total number of localized anyons. However the net magnetic flux seen by the Cooper pair field is 
\begin{equation} 
\int \frac{1}{2\pi} \left(\nabla \times (\bs{a}^f -2 \bs{a}^{b})\right)
\end{equation} 
We expect that the superconductor energetically prefers to have this net flux be zero. Thus we have a ``gauge glass" model with a random sprinkling of $-2\pi$ and $+4\pi$ vortices such that the net vorticity is zero. This model is commonly used in the study of vortex glasses~\cite{fisher1991vortex}. 

It is expected that the AVG, much like the usual vortex glass, is not stable at $T > 0$ due to the creep motion of vortices. Thus there will be no strict finite-$T$ transition into the SC phase for $\rho_s$ sufficiently small. At low but non-zero $T$, the linear resistance will be small but non-zero, and there will be a non-linear response above a small $T$-dependent scale~\cite{fisher1991vortex}. These features of the AVG may explain the gradual decrease of the linear resistance toward zero, and the complicated $I-V$ curves seen in the SC regime in \textit{t}MoTe$_2$.

\section{Transport in a Boltzmann gas of charged anyons}
\label{app:anyongas}

In this section, we provide a semiclassical modeling of the transport properties of a thermal gas of charge-$q$ anyons in the Boltzmann regime. When the number density of anyons is $\delta$, the degeneracy temperature (above which quantum statistics play no role) is on the order of $\delta /m$ where $m$ is the anyon effective mass. Boltzmann transport applies whenever $T \gg \delta /m$. If the anyon gas originates from doping a FQAH state, then $\delta$ is the mobile dopant charge density. An additional upper bound $T \ll \Delta_{\rm Jain}$ is required so that the basic physics responsible for the formation of the anyons is not destroyed by thermal fluctuations. Therefore, we impose the more stringent criterion
\begin{equation}
    \frac{ \delta}{m} \ll T \ll \Delta_{\rm Jain} \,. 
\end{equation}

When the above condition is satisfied, we can write down the semiclassical equations of motion for a single charged anyon in the presence of an applied electric field $\bs{E}$ and magnetic field $\bs{B}$ in the absence of any scattering 
\begin{equation}
    \frac{d \bs{x}}{dt} = \frac{\bs{p}}{m} - \frac{d \bs{p}}{dt} \times \Omega \hat z \,, \quad \frac{d \bs{p}}{dt} = q \left(\bs{E} + \frac{d \bs{x}}{dt} \times \bs{B} \right) \,. 
\end{equation}
In this equation, $\Omega$ is an effective Berry curvature which is generically nonzero in a system with broken time-reversal symmetry. In a classical Hamiltonian system, this Berry curvature arises from a nontrivial Poisson bracket $\{x, y\} = \Omega$ which modifies the symplectic structure on the phase space defined by $(x, y, p_x, p_y)$. Since the precise value of $\Omega$ does not play an important role in our argument, we will leave it as a phenomenological parameter. To linear order in $\bs{E}$ and $\bs{B}$, these equations become 
\begin{equation}
    \frac{d \bs{x}}{dt} = \frac{\bs{p}}{m} - q\bs{E} \times \Omega \hat z \,, \quad \frac{d \bs{p}}{dt} = q \left(\bs{E} + \frac{d \bs{x}}{dt} \times \bs{B} \right) \,. 
\end{equation}
Introducing momentum relaxation at a time scale $\tau$, the second equation becomes 
\begin{equation} 
\frac{d \bs{p}}{dt} = - \frac{\bs{p}}{\tau} + q \left(\bs{E} + \frac{d \bs{x}}{dt} \times \bs{B} \right)
\end{equation}

Solving these equations in the DC limit, we obtain the total conductivity from the anyon gas sector
\begin{equation}
    \sigma = \begin{pmatrix}
        \sigma_0 & \sigma_{\rm anom} + \sigma_{\rm conv}(B) \\ - \sigma_{\rm anom} - \sigma_{\rm conv}(B) & \sigma_0 
    \end{pmatrix} \,, 
\end{equation}
where 
\begin{equation}
    \sigma_0 = \frac{\delta q^2 \tau}{m \left[1 + \left(\omega_c \tau\right)^2\right]} \,, \quad \sigma_{\rm anom} = - \delta q^2 \Omega \,, \quad \sigma_{\rm conv}(B) = \sigma_0 \omega_c \tau \,, \quad \omega_c = qB/m \,. 
\end{equation}
To get the full conductivity, we need to add the contribution from the $\nu = 2/3$ Jain state:
\begin{equation}
    \sigma_c = \begin{pmatrix}
        \sigma_0 & \sigma_{\rm anom} + \sigma_{\rm conv}(B) + 2 e^2/3h \\ - \sigma_{\rm anom} - \sigma_{\rm conv}(B) - 2 e^2/3h & \sigma_0 
    \end{pmatrix} \,. 
\end{equation}
Since $\sigma_0$ is a constant up to $\mathcal{O}(B^2)$ corrections at small $B$, we know that 
\begin{equation}
    \frac{d \sigma^{xy}}{dB}\bigg|_{B=0} = \frac{\delta q^2 \tau}{m} \frac{q \tau}{m} \,. 
\end{equation}
Dividing by the longitudinal conductance then gives 
\begin{equation}
    \frac{1}{\left(\sigma^{xx}\right)^2} \frac{d \sigma^{xy}}{dB}\bigg|_{B=0} = \frac{\delta q^3 \tau^2}{m^2} \frac{m^2}{\delta^2 q^4 \tau^2} = \frac{1}{\delta q} \,. 
\end{equation}
This allows us to infer the mobile carrier density $\delta$.

\end{document}